\documentclass[11pt]{article}
\usepackage{epsfig}
\usepackage{graphics}

\newcommand{\beq}{\begin{equation}}
\newcommand{\eeq}{\end{equation}}

\newcommand{\di}{\displaystyle}

\newcommand{\ga}{\gamma}

\newcommand{\La}{\Lambda}

\newcommand{\si}{\sigma}

\renewcommand{\epsilon}{\varepsilon}

\newcommand{\TeV}{\; \mbox{\rm TeV}}
\newcommand{\GeV}{\; \mbox{\rm GeV}}

\begin{document}
\large

\title{\bf Estimation of charm production cross section in hadronic
interactions at high energies}
\author{G.M. Vereshkov$^{1,2}$\footnote{email: gveresh@ip.rsu.ru}, Yu.F. Novoseltsev$^2$\footnote
{email: YuNovoselt@yandex.ru}}

\date{$^1$Physics Research Institute
of Rostov State University, \protect\\
$^2$Institute for Nuclear Researsh of RAS }

\maketitle

\begin {abstract}
Results of processing experimental data on charm production
in hadron-hadron interactions are presented. The analysis is
carried out within the frame of
phenomenological model of diffraction production and quark
statistics based on additive quark model (AQM). In low energy
region $\sqrt s = 20 - 40$ GeV, the cross sections $ \si _ {pN\to c\bar
cX} (s), \; \si _ {\pi N\to c\bar cX} (s) $ are fitted by logarithmic
function with the parameters connected by relationship of AQM.
 At collider energies 200 $, \; 540, \; 900, \; 1800$ GeV, the values
of $\si _{\bar pp\to c\bar cX} (s)$ were obtained by a quark statistics method
from the data on diffraction dissociation. It is established, that logarithmic
function with universal numerical parameters describes the whole set of
low-energy and high-energy data with high accuracy. The expected values of
cross sections are $\si _{pp\to c\bar cX} = 250\pm 40\,\mu b$ and $355\pm 57\,
\mu b $ at TEVATRON energy $\sqrt {s} = 1.96$ TeV and LHC energy $\sqrt {s}
= 14$ TeV accordingly. Opportunities of use of the obtained results for
calibration of a flux of "prompt" muons in high-energy component of cosmic rays
are discussed.
\end {abstract}

\section{Introduction}

The investigation of charm production processes at overaccelerator
energies is an actual problem of elementary particles and
cosmic rays physics. Nowadays
there are four groups of experimental data:

1) data on total cross sections of charm production in hadron-hadron
interactions $ \si _ {tot} (pN\to c \bar c + X), \; \si _ {tot} (\pi N\to c
\bar c + X) $ in the region of low energies $ \sqrt {s} =20 \; - \; 40 \GeV$;

2) data on total cross sections of charm production in a photon - hadron
interactions $ \si _ {tot} (\ga N\to c \bar c + X) $ at low energies $ \sqrt
{s} =6 \; - \; 20\, \GeV $ and - at virtualities of a photon up to $Q^2=4\,
\GeV^2 $ - in the region of high energies $ \sqrt {s} =160 \; - \; 240\, \GeV
$;

3) data on production of $J/\Psi, \; \Psi (2S) $ particles in $p\bar
p$-interactions at $ \sqrt {s} =1800\GeV $ \cite {4}, \cite {5}.

4) data on differential cross section of charm production in $p\bar p-$
interaction in the central area at $ \sqrt {s} =1960\GeV $ and transverse
momentum  $p_T > 5.5\GeV $ \cite {5a}.

The first two groups of the data and the appropriate
review of the literature contain, for example, in Ref.  \cite {2}.
Only the first group of the data has the status of a direct
information on charm production in hadron-hadron interactions
 Three other groups differ from the first one by initial or
final states, therefore the joint model independent processing
 of all three groups of the data is impossible. Accordingly,
comparison of the theory to experiment is at a loss also: fixing
of parameters of theoretical model by its comparison to the one group
of high-energy data results in catastrophic disagreements
of the same model with other group of the data. In the scientific
literature this situation is characterized as "an unsolved problem"
\ \cite {3}.

In the present work an estimation of the total charm production cross
section is offered. The estimation is based on processing of
collider data on diffraction dissociation with use of phenomenological
model of diffraction charm production and quark statistics.
 The result reduces to simple formula with well enough certain
numerical parameters:
\beq
\begin {array} {c}
\di \si _ {p\bar p/pp\to c \bar c + X} (s) =C _ {p\bar p/pp} \ln\frac {s} {s_0},
\\ [5mm]
\di C _ {p\bar p/pp} =26.78\pm 1.44\, \mu b, \qquad \sqrt {s_0} =18.23\pm
0.33\GeV.
\end {array}
\label {fin}
\eeq
For TEVATRON energy $ \sqrt {s} =1800\GeV $ our estimation gives $\si _ {\bar
pp\to \bar cc+X} \simeq 246\pm 30\,\mu b $; for LHC energy $\sqrt {s}
=14000\GeV, \; \si _ {pp\to c\bar c+X} =\simeq 356\pm 35\,\mu b $.

In cosmic ray (CR) physics charmed particles, arising in interaction of primary
particle with an atmosphere, are sources of so called "prompt" \ high-energy
muons, carrying away energy from Extensive Air Shower (EAS). A well known
problem about the nature of a break ("the knee") in the cosmic rays energy
spectrum at $E_{k (Lab)} \approx 3\times 10^3\TeV \; \; (\sqrt {s_k} \approx
2.5\TeV)$ can not be solved without the quantitative information on the
contribution of charmed particles in the total muon flux. Obtaining such
information is a complex problem which is solved by joint processing the
accelerating data (LHC data will be added to them in the coming years) and
results which should be received at CR facilities. These are accelerating data
on total and inclusive cross sections of hadronic interactions (including cross
sections of charm production) and CR data on the flux of muons with energies
$E_{\mu (Lab)} \ge 70\TeV$.

In conditions of incompleteness of accelerating data, the measurements of
"prompt"\ muons fluxes at energies $E_{\mu} \ge 70\TeV $ gain special interest
- their results are capable to give the preliminary information about a
character of hadronic interactions in the range of ultrahigh energies. For
planning and preparation of the experiment, the forecast of expected value of
charm production cross section is necessary, provided that laws of this process
are extrapolated from accelerating energy region. Our estimation represents one
of the variants of such forecast which can be used for calibration of the flux
of ultrahigh energy CR muons.

\section {Diffraction model of charm production and additive quark model}

The offered method of the estimation of charm production cross
section is based on the usage of additional set of experimental data.
 The question is cross sections of proton  diffraction dissociation
in $p\bar p $ - interactions at SPS CERN energies $\sqrt {s} = 200\GeV, \;
900\GeV$ \cite {10} and TEVATRON ones $\sqrt {s} = 546\GeV, \; 1800\GeV$ \cite
{11}. According to these data, the diffraction dissociation cross section
depends logarithmically on the energy:
\beq
\di \si _ {DD} (p\bar p\to X) =C _ {DD} \ln\frac {s} {s_0}.
\label {1}
\eeq

The basic assumptions consist in the following:\\
1) charm
production occurs in process of diffraction dissociation;\\
2) charm production cross section is extracted from total
cross section (\ref {1}) by quark statistics rules
obtained from additive quark model:
\beq
\di \si _ {tot} (p\bar p\to c \bar c + X) \simeq k _ {c \bar c} \times
\si _ {DD} (p\bar p\to X), \qquad k _ {c \bar c} \simeq 0.025\pm 0.004.
\label {2}
\eeq

These assumptions, from our point of view, have general enough
phenomenological character.
Formulas (\ref {1}), (\ref {2}) have the status of the asymptotic
ones. They are fulfilled more precisely, the greater is the energy
of interaction. The value of $k_{c \bar c} \simeq 0.025$,
presented in (\ref {2}), is obtained from the relationships in
Ref. \cite {12}:
\beq
\di u\bar u \;: \; d\bar d \;: \; s\bar s \;: \; c\bar c = 1
\;: \; 1 \;: \; 0.38\pm0.07 \;: \; 0.06\pm0.01,
\label {4}
\eeq
here the symbol $q\bar q $ designates the number of secondary hadrons
containing quark $q $ or antiquark $\bar q$. Existence of similar relationships
is predicted by additive quark model (AQM) \cite {13}. Direct check of these
relationships is realized in experiments, in which the composition of secondary
particles is supervised. A condition of inclusion of charmed particles in quark
statistics is an absence of energy restrictions on process of their production.
This condition is obviously fulfilled for charmed particles produced in the
diffraction region. All numbers appearing in (\ref {4}), except the one
concerning to $c-$ quarks, are obtained by comparison of AQM to the experiment
at $\sqrt {s} = 540\GeV$ \cite {14}. An estimation of the suppression factor of
charmed hadrons $\lambda_c / \lambda_s \approx 0.15$ is based on characteristic
assumption for AQM: for $q=s, \; c, \; b$
\[
\di \lambda_q \sim \sum_i \frac {2J_i+1} {M_i^2},
\]
where the summation is over all particles and resonances with spins $J $ and
masses $M $, containing a quark $q $ or an antiquark $ \bar q $.
Each of the statements reflected in formulas
(\ref {1}) - (\ref {4}) was repeatedly
discussed in scientific literature. We shall note, that typical
quantitative accuracy of the phenomenological models
based on these statements is 10 - 20 \%.

First of all we shall find out the opportunities of AQM and logarithmic
dependence for cross sections in the description of direct experimental data on
charm production in $pN $ and $\pi N$ interactions at energies $ \sqrt {s} =20
\;-\; 40\GeV $. These two hypotheses are reduced to formulas
\beq
\di \si _ {pN\to c \bar c + X} (s) = C_{pN} \ln\frac {s}{s_0},
\qquad \si _{\pi N\to c \bar c + X} (s) = \frac {2} { 3 } \si _ {pN\to c \bar c + X} (3s/2),
\label {5}
\eeq
containing only two free parameters. Numerical values of these
parameters are defined by joint fitting of $pN$ and $\pi N$ data:
\beq
\di C _ {pN} =28.84\pm 2.10\, \mu b, \qquad \sqrt {s_0} =18.51\pm 0.36\GeV,
\qquad \chi^2=0.89.
\label {6}
\eeq

\begin{figure}[h]
\epsfxsize=0.7\textwidth \centerline{\includegraphics{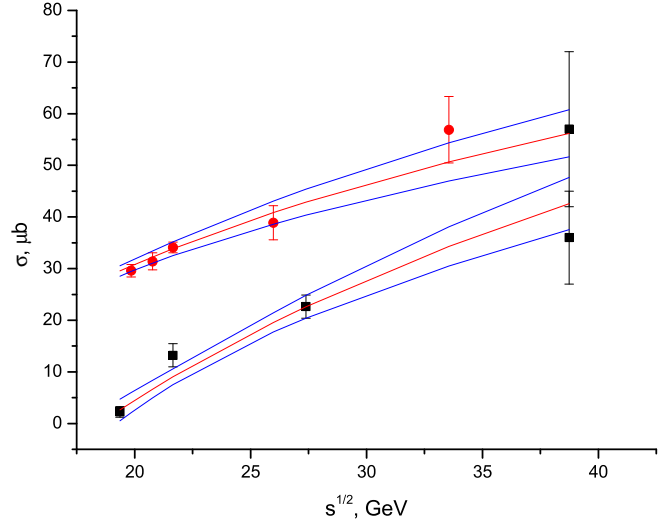}}
\caption{Accelerator data on charm production cross section in pN  and $\pi N$
interactions. $\pi N$ data are shifted up for clearness.}
\end{figure}

Results of fitting are shown in Fig. 1 ($\pi N$ data are shifted up on $20\ \mu
b$ for clearness). As it follows from Fig. 1, the phenomenological diffraction
model of charm production describes well the direct low-energy data. However
extrapolation of these results in the region of high energies $\sqrt {s} = 1000
- 3000\GeV$ has no substantiation, certainly.

As it was already marked in the beginning of the Section,
there is an opportunity to involve the additional experimental
information in the range of SPS - TEVATRON energies.
Cross sections of diffraction dissociation
measured at these colliders have the following values:
\[
\begin {array} {lllll}
\di \sqrt {s}, \GeV & \ \ \ 200 & \ \ \ 546 & \ \ \ 900 & \ \ \ 1800
\\ [7mm]
\di \si _ {DD},\ mb  & 4.8\pm 0.5\qquad & 7.89\pm 0.33\qquad &
7.8\pm 0.5\qquad & 9.46\pm 0.44
\end {array}
\]
It is easy to verify that the given numbers are fitted in
logarithmic dependence. Calculation of values of charm production
cross sections from the formula (\ref {2}), taking into account
relationships of AQM ( \ref {4}), gives:
\[
\begin {array} {lllll}
\di \sqrt {s}, \GeV & \ \ \ 200 & \ \ \  546 & \ \ \  900 & \ \ \ 1800
\\ [7mm]
\di \si _ {p\bar p\to c \bar c + X}, \,\mu b  &
 120\pm 21 \qquad & 197\pm 32 \qquad &
 195\pm 32\qquad & 236\pm 38
\end {array}
\]

The obtained values of charm production cross sections have
the status of model dependent processing collider data.

\begin{figure}[h]
\epsfxsize=0.7\textwidth
\centerline{\includegraphics{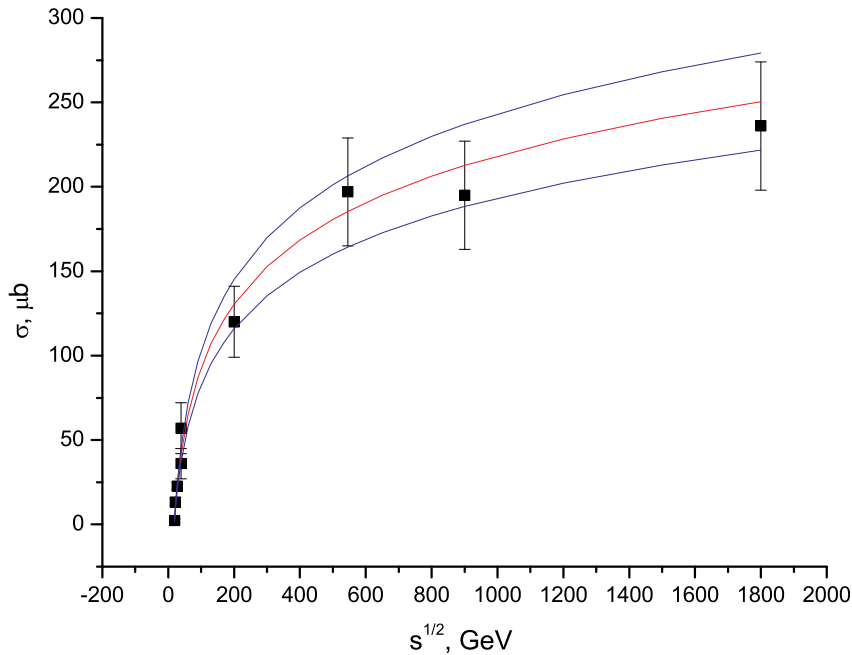}}
\caption{Joint fit of low-energy and high-energy data of charm production.}
\end{figure}

In Fig. 2 these results
are presented together with the data of direct measurements
in the region of low energies. The result of fitting the total data
set is:
\beq
\di C_{p\bar p} = 26.78\pm 1.44\ \mu b, \qquad \sqrt {s_0} =18.23\pm 0.33 \GeV,
\qquad \chi^2 = 0.98.
\label {7}
\eeq
As it is evident from the foregoing, values of the parameters
(\ref {6}) and (\ref {7}) coincide
within the limits of statistical errors. It means the charm
production cross section in hadronic interactions is described
by universal logarithmic dependence covering both low-energy and
high-energy experimental data.

\section {Discussion of results and conclusions}

Only the accuracy level is surprising with which
low-energy (\ref {6}) and total (\ref {7}) parameters
of logarithmic function coincide. With regard to ideas of
phenomenological models of diffraction production and quark
statistics, they were repeatedly tested at processing most
various experimental data.
These models give reliable quantitative predictions at the
accuracy level 10 - 20 \% after fixing several parameters by
experimental data. Therefore we consider possible to use
the results (\ref {7}) for forecasting cross section of charm
production in $pp/p\bar p$ interactions in overaccelerating
energy range.

\begin{figure}[h]
\epsfxsize=0.7\textwidth
\centerline{\includegraphics{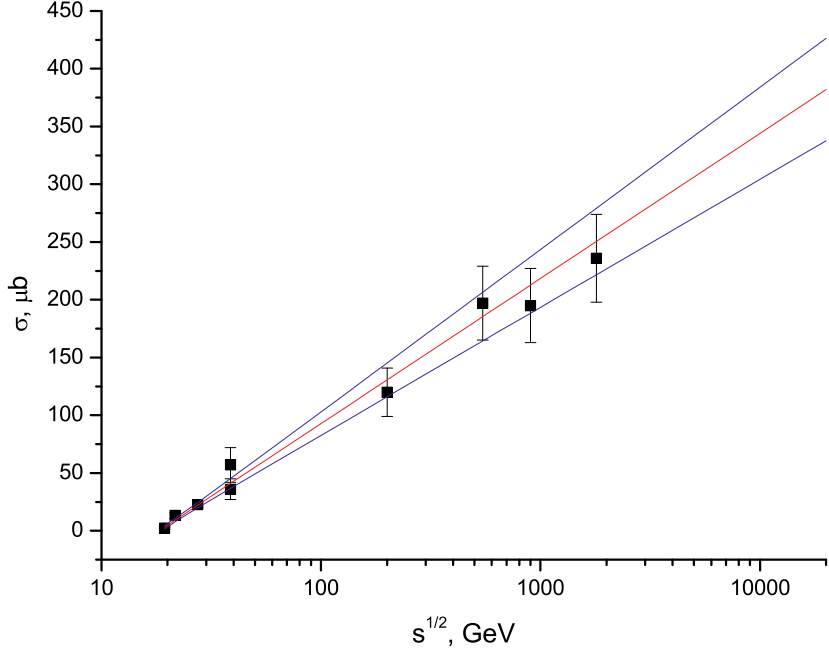}}
\caption{Extrapolation of the charm production accelerator
data fit to overaccelerating energy range.}
\end{figure}

In Fig. 3 the dependence of charm production cross section on energy is
presented. In Figure is shown the dependence (\ref {fin}) with parameters (\ref
{7}) extrapolated in overaccelerating energy range. The confidence strip of the
forecast corresponds to 90 \% CL. The expected value of cross section is
$\si_{pp\to c\bar c+X} \simeq 270 - 300\, \mu b $ in the energy region $\sqrt
{s} \simeq 3000 - 5000\GeV $ and $\si _ {pp\to c\bar c+X} \simeq 355\pm 35\,\mu
b$ at LHC energy $\sqrt {s} = 14000\GeV$.

Using the quark statistics, the total cross section of charm production can be
presented as the sum of inclusive cross sections. Appropriate relationships
between cross sections \cite {2} are checked experimentally up in the region $
\sqrt {s} = 20 - 40\GeV$:
\[
\di \si (D_s\bar D_s)/\si (D^0\bar D^0+D ^ +\bar D ^-)\simeq 0.2, \qquad
\si (\La_c\bar D)/\si (D^0\bar D^0+D ^ +\bar D ^-)\simeq 0.3
\]
As it was shown above, a hypothesis of scale invariance of total
cross section parameters is experimentally motivated; therefore
we expand this hypothesis on the parameters of inclusive cross
sections. In this case we have in laboratory system
\beq
\begin {array} {c}
\di \si _ {D\bar D} (E) \approx 154.1+17.86\ln E \ \mu b,
\\ [6mm]
\di \si _ {D_s\bar D_s} (E) \approx 30.8+3.55\ln E \ \mu b,
\\ [6mm]
\di \si _ {\La_c\bar D} (E) \approx 46.2+5.33\ln E \ \mu b.
\end {array}
\label {si}
\eeq
where E is in PeV ($1 $ PeV$ = 10^3\TeV$).
 A technique of use of similar type formulas
for calculation of an expected flux of "prompt" \ high-energy muons,
accompanying EAS, is described in Ref. \cite {16}. Detection of such
muons, in essence, is check of numerical values of factors
appearing in formulas (\ref {si}).

Now the experiment on recording very high energy (VHE) muons
($E_\mu \ge $ 70 TeV) is carried out at Baksan Underground
Scintillation Telescope (BUST) \cite {B}. Preliminary results
of this experiment, expected in the near future, will show as far as
the offered estimation of charm production cross section is
adequate.
It should be noted that limiting energy of TEVATRON $\sqrt {s_{max}}
=1.96\TeV$ is very close to the energy of the break in CR energy
spectrum $ \sqrt {s _ {knee}} \approx 2.5\TeV $. If total cross sections of
charm production would be measured at TEVATRON, their small extrapolation would
allow to obtain the information necessary for calibration of the flux of VHE
muons. In this case, logic and interpretation of the experiment on recording CR
muons would get at once a rigid character.

It should be pointed out, there are three possible (and alternative)
variants of results, each of which represents the certain interest.

1) The measured flux of CR muons within the limits of measurement
errors coincides with calibration flux calculated according to the
estimation of charm production cross section (\ref {fin}). In this
case the formula (\ref {fin}) can be used for testing microscopic models.

2) The measured flux of CR muons will exceed noticeably the
calibration value, but the appropriate cross sections of
charm production at energies in the region of a break of CR
energy spectrum
will stay essentially smaller $10\ mb$. In this case
it will be necessary to recognize, that either quark statistics rules,
in that kind in which they were used in the present work,
become incorrect in the area of high energies, or
the new mechanism of charm production,
distinct from the diffraction one, is included at these energies.

3) The measured flux of CR muons will correspond formally
to charm production cross section exceeding $10\ mb$. This result will
mean that VHE muons carry away the energy from EAS
forming an observable break of CR spectrum (see, for example,
\cite {P}). However, it is necessary to note, that sources of
such big numbers of VHE muons, most likely, are not reduced to charmed
particles - a New physics will required for interpretation of such
effect.
Certainly for the formulation of new ideas, the data on the flux
of CR muons should be considered together with all
other accelerating and CR data.

\end{document}